\documentclass[useAMS,usenatbib,]{mn2e}

 \usepackage{graphicx}
\usepackage{amsmath,amsfonts,amssymb}
\usepackage{color}

\title{Signatures of pulsars in the light curves of newly formed supernova remnants}

\author[K. Kotera, E. S. Phinney, and A. V. Olinto]{K. Kotera$^{1,2}$\thanks{E-mail:
kotera@iap.fr }, E. S. Phinney$^{2}$, and A. V. Olinto$^{3}$\\
$^{1}$Institut d'Astrophysique de Paris, UMR7095 - CNRS, Universit\'e Pierre \& Marie Curie, \\98 bis boulevard Arago F-75014 Paris, France\\
$^{2}$California Institute of Technology, Mailcode 350-17, 1200 E California Blvd, Pasadena CA, 91125\\
$^{3}$Department of Astronomy \& Astrophysics, Enrico Fermi Institute, \\and Kavli Institute for Cosmological Physics, The
  University of Chicago, Chicago, Illinois 60637, USA.}
\begin{document}

\date{\today}

\pagerange{\pageref{firstpage}--\pageref{lastpage}} \pubyear{2012å}

\maketitle

\label{firstpage}

\begin{abstract}
We explore the effect of pulsars, in particular those born with millisecond periods, on their surrounding supernova ejectas. While they spin down, fast-spinning pulsars release their tremendous rotational energy in the form of a relativistic magnetized wind that can affect the dynamics and luminosity of the supernova. We estimate the thermal and non thermal radiations expected from these specific objects, concentrating at times a few years after the onset of the explosion. We find that the bolometric light curves present a high luminosity plateau (that can reach $10^{43-44}\,$erg/s) over a few years. An equally bright TeV gamma-ray emission, and a milder X-ray peak (of order $10^{40-42}\,$erg/s) could also appear a few months to a few years after the explosion, as the pulsar wind nebula emerges, depending on the injection parameters. The observations of these signatures by following the emission of a large number of supernovae could have important implications for the understanding of core-collapse supernovae and reveal the nature of the remnant compact object.
\end{abstract}


\section{Introduction}

Core-collapse supernovae are triggered by the collapse and explosion of massive stars, and lead to the formation of black holes or neutron stars (see, e.g., \citealp{Woosley02}). In particular, pulsars (highly magnetized, fast rotating neutron stars) are believed to be commonly produced in such events. The observed light curves of core-collapse supernovae present a wide variety of shapes, durations, and luminosities, that many studies have endeavored to model, considering the progenitor mass, explosion energy, radioactive nucleosyntheis, and radiation transfer mechanisms in the ejecta (e.g., \citealp{Hamuy03,Utrobin08,Baklanov05,Kasen09}). 

While they spin down, pulsars release their rotational energy in the form of a relativistic magnetized wind. The effects of a central pulsar on the early supernova dynamics and luminosity is usually neglected, as the energy supplied by the star is negligible compared to the explosion energy, for the bulk of their population. Some pioneering works have however sketched these effects \cite{Gaffet77b,Gaffet77a,Pacini73,Bandiera84,Reynolds84}, notably in the case of SN 1987A \citep{McCray87,Xu88}. More recently, \cite{Kasen10,Dessart12} discussed that magnetars, a sub-class of pulsars born with extremely high dipole magnetic fields of order $B\sim 10^{14-15}\,$G and millisecond spin periods, could deposit their rotational energy into the surrounding supernova ejecta in a few days. This mechanism would brighten considerably the supernova, and could provide and explanation to the observed superluminous supernovae \citep{Quimby12}.

In this paper, we explore the effects of mildly magnetized pulsars born with millisecond periods (such as the Crab pulsar at birth) on the light curves of the early supernovae ejecta. Such objects are expected to inject their tremendous rotational energy in the supernovae ejecta, but over longer times compared to magnetars (of order of a few years). Indeed, the spin-down and thus the timescale for rotational energy deposition is governed by the magnetization of the star. 

We estimate the thermal and non thermal radiations expected from these specific objects, concentrating at times of a few years after the onset of the explosion. We find that the bolometric light curves present a high luminosity plateau (that can reach $10^{43-44}\,$erg/s) over a few years, and that an equally bright TeV gamma-ray emission could also appear after a few months to a few years, from the acceleration of particles in the pulsar wind, depending on the injection parameters. A milder associated X-ray peak (of luminosity $10^{40-42}\,$erg/s) could also be produced around the same time. The observations of these signatures by the following up of a large number of supernovae could have important implications for the understanding of core-collapse supernovae and reveal the nature of the remnant compact object.

These objects also present the ideal combination of parameters for successful production of ultrahigh energy cosmic rays (UHECRs, see \citealp{Blasi00,Fang12}). The observation of such supernovae could thus be a further argument in favor of millisecond pulsars as sources of UHECRs, and a potential signature of an ongoing UHECR production. 

We first give, in Section~\ref{section:properties}, the list of quantities necessary for this analysis in the regimes of interest for the ejecta: optically thin or thick, and present a scheme of the early interaction between the pulsar wind and the supernova ejecta. In Section~\ref{section:lightcurves}, we calculate the bolometric, thermal, and non thermal light curves of our peculiar supernovae. In Section~\ref{section:discussion}, we briefly discuss available observations, and the implications for UHECR production.

\section{Supernova ejecta hosting a millisecond pulsar: properties}\label{section:properties}

We note $M_{\rm ej}$ and $E_{\rm ej}$ as the mass and initial energy of the supernova ejecta. The pulsar has an inertial momentum $I$, radius $R_*$, initial rotation velocity $\Omega_{\rm i}$ (corresponding initial period $P_{\rm i}=2\pi/\Omega_{\rm i}$), and dipole magnetic field $B$.
Numerical quantities are noted $Q_x\equiv Q/10^x$ in cgs units, unless specified otherwise.

\subsection{Timescales}\label{subsection:timescales}

\begin{figure}
\begin{center}
\includegraphics[width=\columnwidth]{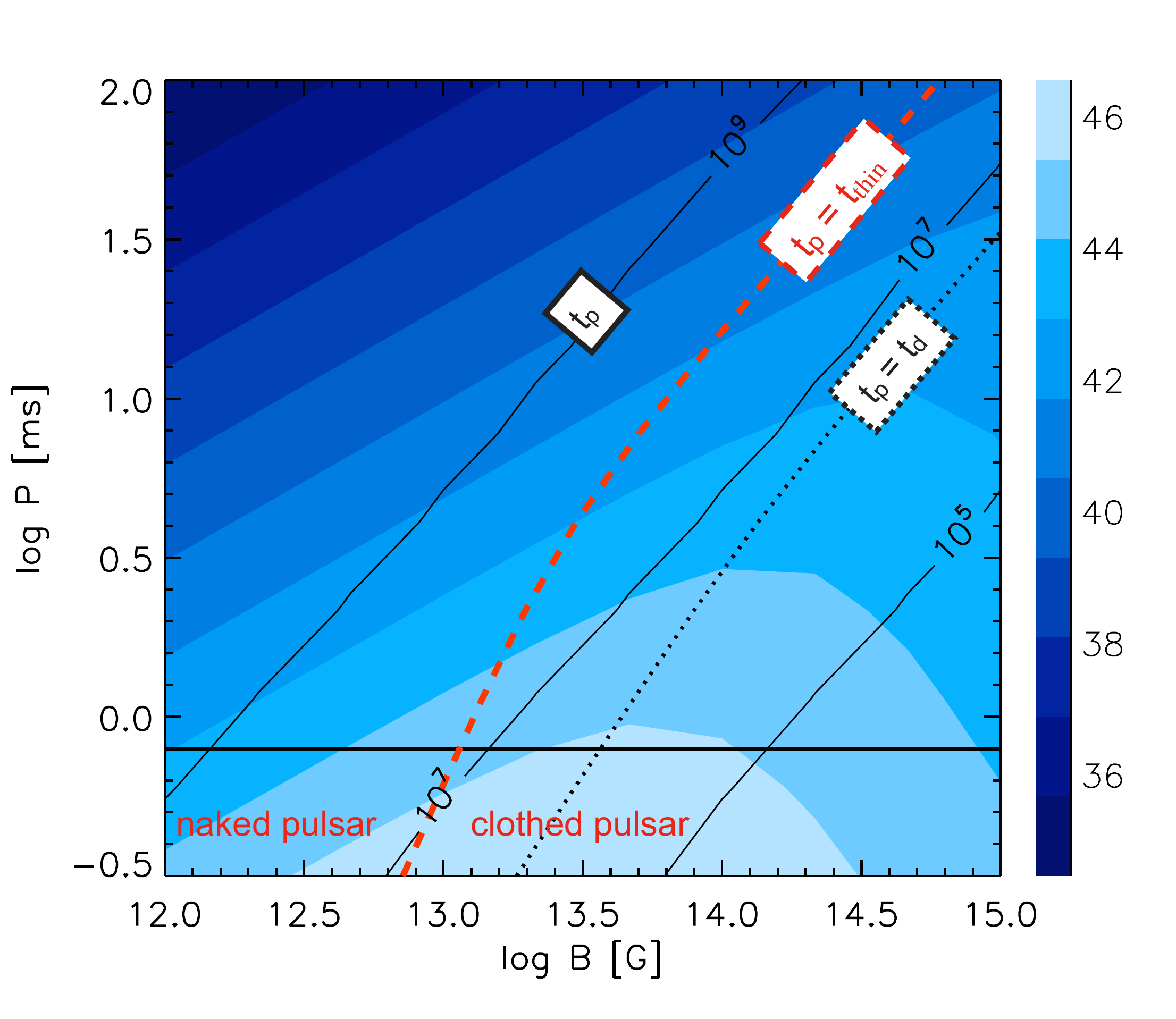} 
\caption{{Contour plot of the bolometric luminosity of  supernova+pulsar wind nebula systems at 1\,yr after explosion (the fraction of wind energy converted into radiation, as defined in Section~\ref{section:lightcurves} is set to $\eta_\gamma=1$), as a function of initial period $P$ and magnetic field $B$. The various regimes for radiative emissions described in Section~\ref{subsection:timescales} are represented. The solid lines indicate pulsar spin-down timescale in seconds (Eq.~\ref{eq:tp}). The red dashed lines represent the pulsar population for which $t_{\rm p}=t_{\rm thin}$, and separate naked and clothed pulsars (see text). The dotted lines represent $t_{\rm p}=t_{\rm d}$.}}\label{fig:SNLum}
\end{center}
\end{figure}

For an ordinary core-collapse supernova, the ejecta expands into the circumstellar medium at a characteristic final velocity 
\begin{equation}
v_{\rm ej}=v_{\rm SN}=\left(2\frac{E_{\rm ej}}{M_{\rm ej}}\right)^{1/2}\sim 4.5\times 10^8\,\mbox{cm\,s}^{-1}\,E_{\rm ej,51}^{1/2}M_{\rm ej,5}^{-1/2} \ ,
\end{equation}
where $M_{\rm ej,5}\equiv M_{\rm ej}/5\,M_\odot$. 
After a few expansion timescales $t_{\rm ex} = R_{\rm ej,i}/v_{\rm SN}$, where $R_{\rm ej,i}$ is the radius of the star that led to the explosion, the ejecta enters into a stage of homologous expansion where its size scales as $R=v_{\rm ej}t$ and its internal energy as $E_{\rm int}(t) \sim (E_{\rm ej}/2)(t_{\rm ex}/t)$.\\

The ejecta is first optically thick to electron scattering. Noting $\kappa$ and $\rho$ the opacity and density of the supernova envelope, one can estimate the optical depth of the ejecta:
$\tau = R\kappa\rho $.
Assuming a constant central supernova density profile (see \citealp{Matzner99} and \citealp{Chevalier05} for more detailed modeling of the interior structure of supernovae) $\rho = 3M_{\rm ej}/(4\pi R^3)$, one can define the effective diffusion time (for thermal photons to cross the ejecta):
\begin{eqnarray}
t_{\rm d} &\equiv& \left(\frac{M_{\rm ej}\kappa}{4\pi v_{\rm ej}c}\right)^{1/2} \\
&\sim& 1.6\times 10^{6}\,{\rm s}\,M_{\rm ej,5}^{1/2}\kappa_{0.2}^{1/2}\left(\frac{v_{\rm ej}}{2\times 10^{9}\,{\rm cm\,s}^{-1}}\right)^{-1/2}\ ,
\end{eqnarray}
with the opacity to electron scattering defined as $\kappa_{0.2}\equiv \kappa/(0.2\, {\rm g^{-1}\,cm}^{2}$) for thermal photons.
This sets the timescale of the supernova light curve, under the assumption that the opacity remains constant throughout the ejecta (no ionization effect), and in the absence of pulsar or $^{56}$Ni heating. For more detailed computation of the these timescales, see, e.g., \cite{Kasen09}. \\

As the ejecta expands, it reaches a time $t_{\rm thin}$ when it becomes optically thin to electron scattering, for thermal photons ($\tau=1$):
\begin{equation}
t_{\rm thin} = \left(\frac{3 M_{\rm ej}\kappa}{4\pi v_{\rm ej}^2}\right)^{1/2} \sim 1.9\times 10^7 \,{\rm s}\left(\frac{v_{\rm ej}}{2\times 10^{9}\,{\rm cm\,s}^{-1}}\right)\ .
\end{equation}
For the numerical estimates of $v_{\rm ej}$, we are using the final velocity of the ejecta after its modification by the shock at the interface between the pulsar wind and the initial ejecta, for $E_{\rm ej,51}$, $M_{\rm ej,5}$, and $P_{\rm i}=10^{-3}\,$s (see Eq.~\ref{eq:vej} in Section~\ref{subsection:velocities}).

The pulsar spins down by electromagnetic energy losses that is transferred to the surrounding environment. 
The deposition of this energy happens over the spin-down timescale of the pulsar \citep{Shapiro83}:
\begin{equation}\label{eq:tp}
t_{\rm p} = \frac{9Ic^3}{2B^2R_*^6\Omega_{\rm i}^2} \sim 3.1\times 10^{7}\,{\rm s}\,I_{45}B_{13}^{-2}R_{*,6}^{-6}P_{\rm i,-3}^2\ .
\end{equation}

We will consider two regimes for the calculation of radiative emissions from the ejecta: optically thin ($t>t_{\rm thin}$), and optically thick ($t<t_{\rm thin}$) for thermal photons. The deposition of pulsar rotational energy will have different effects on the supernova radiative emissions according to the optical depth of the ejecta at time $t_{\rm p}$. Figure~\ref{fig:SNLum} pictures these various regimes. The red dashed lines represent the pulsar population for which $t_{\rm p}=t_{\rm thin}$: on its left-hand side, most of the rotational energy of the pulsar is injected when the supernova ejecta is optically thin to electron scattering (the pulsar is {\it naked}). On the right-hand side of the red dashed line, the pulsar energy can enhance the luminosity of the supernova, as it is injected while the ejecta is still optically thick (the pulsar is {\it clothed}).

\begin{figure}
\begin{center}
\includegraphics[width=\columnwidth]{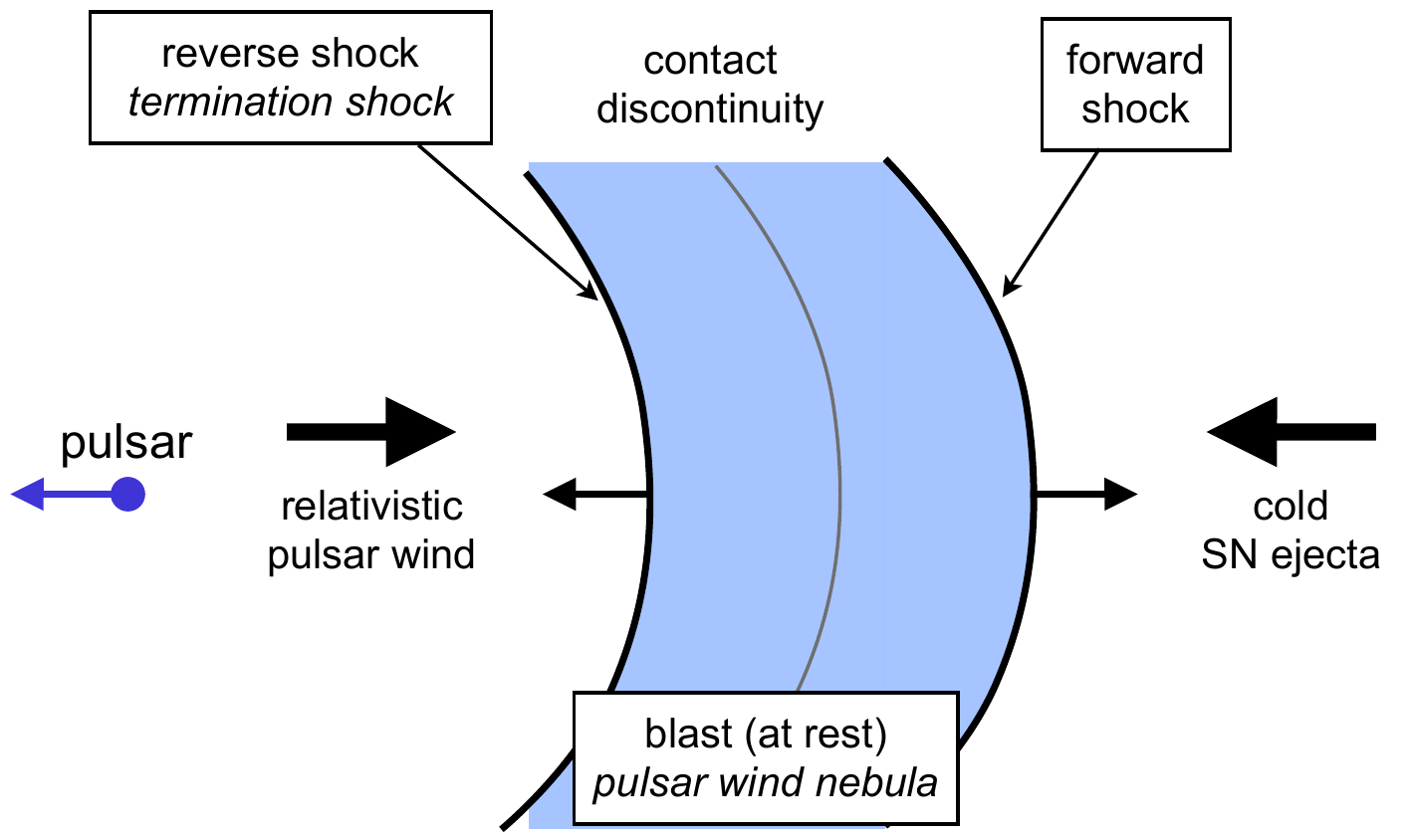} 
\caption{Scheme of the structures created by the interaction between the pulsar wind and the SN ejecta in the blast rest mass frame.}\label{fig:shocks}
\end{center}
\end{figure}

\subsection{Characteristics of the supernova ejecta and of the embedded pulsar wind nebula}\label{subsection:velocities}

The interaction between the pulsar wind and the supernova ejecta leads to the formation of the following structures, illustrated in Fig.~\ref{fig:shocks}: a forward shock at the interface of the shocked and unshocked ejecta, and a reverse shock at the interface between the shocked and unshocked wind (commonly called ``termination shock''). The shocked material between the forward and the reverse shock constitutes the pulsar wind nebula (PWN, e.g., \citealp{Chevalier77,Chevalier92,Gaensler05}).

\begin{figure}
\begin{center}
\includegraphics[width=0.49\textwidth]{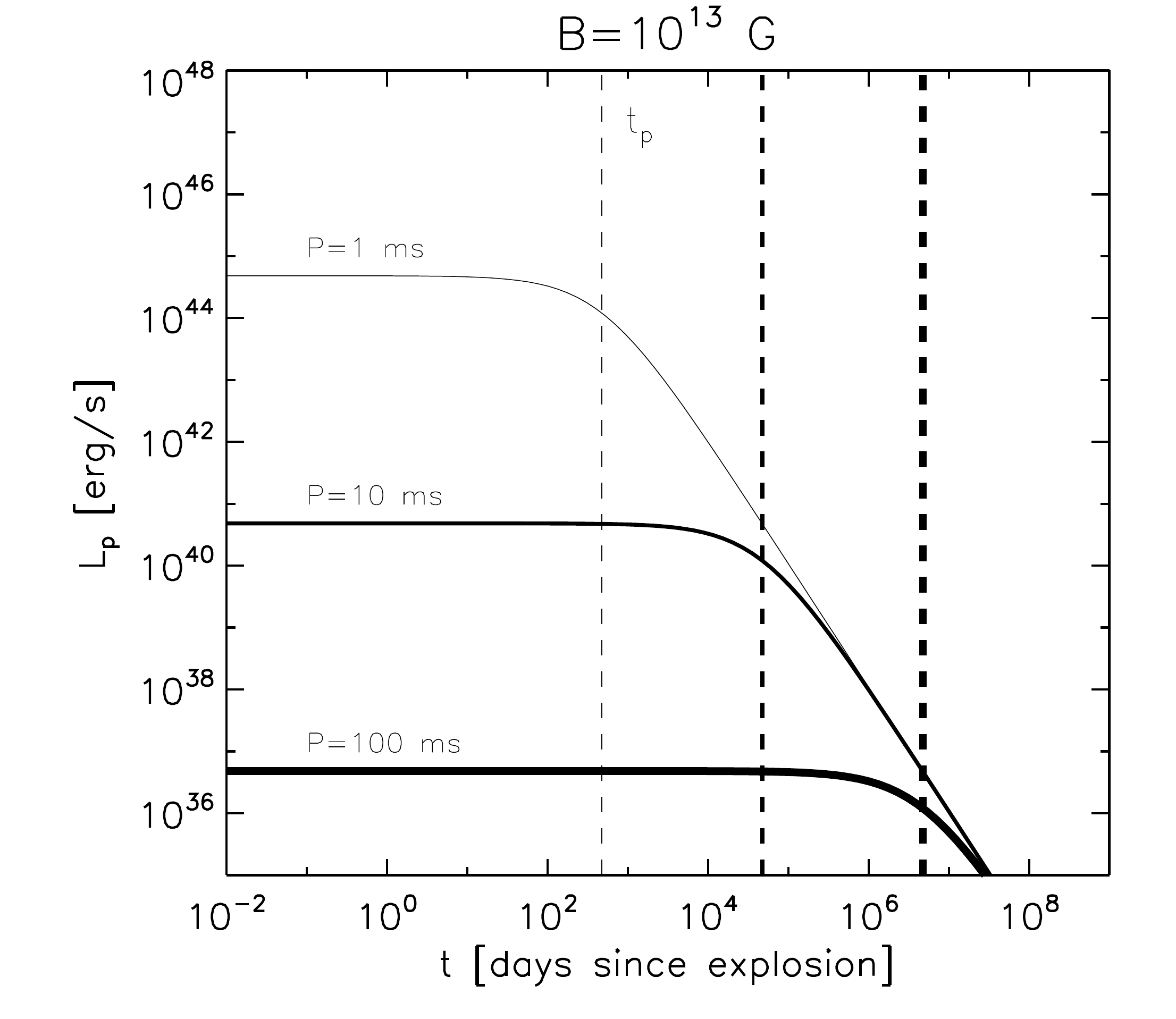} 
\caption{Evolution of the pulsar luminosity $L_{\rm p}$ as a function of time, for magnetic dipole spin-down. The pulsar has a dipole magnetic field of $B=10^{13}\,$G, and period $P_{\rm i}=1, 10, 100\,$ms (increasing thickness). The vertical dashed lines correspond to the spin-down timescale $t_{\rm p}$ for each intial spin period.}\label{fig:Lp_t}
\end{center}
\end{figure}

The pulsar wind carries a total energy:
\begin{equation}
E_{\rm p} = \frac{I\Omega_{\rm i}^2}{2} \sim 1.9\times 10^{52}\,{\rm erg}\, I_{45}P_{\rm i,-3}^2\ ,
\end{equation}
and injects a luminosity (Shapiro \& Teukolsky 1983)
\begin{equation}
L_{\rm p}(t) = \frac{E_{\rm p}}{t_{\rm p}}\frac{1}{(1+t/t_{\rm p})^2}\ .
\end{equation}
into the cold supernova ejecta. The evolution of the pulsar luminosity over time, for magnetic dipole spin-down, is represented in Fig.~\ref{fig:Lp_t}. 

The characteristic velocity of the ejecta is not affected by the pulsar wind nebula expansion if $E_{\rm p}\ll E_{\rm ej}$. However, if the pulsar input energy overwhelms the initial ejecta energy $E_{\rm p}\gg E_{\rm ej}$, the ejecta is swept up into the shell at a final shell velocity $v_{\rm f}=(2E_{\rm p}/M_{\rm ej})^{1/2}$ \citep{Chevalier05}.
Taking into account these two extreme cases, one can estimate the characteristic ejecta velocity as
\begin{equation}\label{eq:vej}
v_{\rm ej} = v_{\rm SN}(1+E_{\rm p}/E_{\rm SN})^{1/2}\ .
\end{equation}

For $E_{\rm p}\ll E_{\rm ej}$, the evolution of the pulsar wind nebula takes place in the central part of the SN ejecta, where the density profile is nearly flat, with $\rho\propto t^{-3}(r/t)^{-m}$. We will assume here that $m=0$. For times $t\le t_{\rm p}$ where $L_{\rm p}\sim E_{\rm p}/t_{\rm p}$, the radius of the pulsar wind nebula can then be expressed~\citep{Chevalier77}
\begin{eqnarray}\label{eq:Rpwn}
R_{\rm PWN} &\sim& \left( \frac{125}{99}\frac{v_{\rm ej}^3E_{\rm p}}{M_{\rm ej}t_{\rm p}}\right)^{1/5}\,t^{6/5}, \mbox{ for } t\le t_{\rm p}, \, E_{\rm p}\ll E_{\rm ej}
\end{eqnarray}

Beyond the characteristic velocity $v_{\rm SN}$, the density profile of the ejecta steepens considerably, reaching spectral indices $b\gtrsim 5$ (e.g., \citealp{Matzner99}). For $E_{\rm p}\gg E_{\rm ej}$, the pulsar wind nebula expands past this inflection point and its size depends on whether the swept-up shell breaks up by Rayleigh-Taylor instabilities. \cite{Chevalier05} discusses that if the shell does not break up, the expansion is determined by the acceleration of a shell of fixed mass, thus,  $\mbox{for } t\le t_{\rm p},\, E_{\rm p}\gg E_{\rm ej}, \,\mbox{and no shell disruption} $
\begin{eqnarray}
R_{\rm PWN} &=&  \left( \frac{8}{15}\frac{E_{\rm p}}{M_{\rm ej}t_{\rm p}}\right)^{1/2}\,t^{3/2} \label{eq:Rpwn_noshelldisr}\\
&\sim &2.2\times 10^{16}\,{\rm cm}\,E_{\rm p,52}^{1/2}M_{\rm ej,5}^{-1/2}t_{\rm p,yr} \quad \mbox{for } t=t_{\rm p}.
\end{eqnarray}
Otherwise, the evolution of the nebula is set by pressure equilibrium, and $R_{\rm PWN}\propto t^{(6-b)/(5-b)}$ (for $t<t_{\rm p}, \, E_{\rm p}\gg E_{\rm ej}$). In the following, because the fate of the shell is unclear at this stage, we will use Eq.~(\ref{eq:Rpwn_noshelldisr}) as an illustration. 

For $t>t_{\rm p}$, $L_{\rm p}$ drops, and the swept-up material tends towards free expansion. One can roughly assume the relation
\begin{equation}\label{eq:Rpwn_tp}
R_{\rm PWN}(t>t_{\rm p}) = R_{\rm PWN}(t_{\rm p})\frac{t}{t_{\rm p}}\ ,
\end{equation}
where $R_{\rm PWN}(t_{\rm p})$ is the size of the pulsar wind nebula in Eqs.~(\ref{eq:Rpwn},\ref{eq:Rpwn_noshelldisr}).
More detailed modelings of the dynamical evolution of pulsar-driven supernova remnants can be found in \cite{Reynolds84}.

\begin{figure}
\begin{center}
\includegraphics[width=\columnwidth]{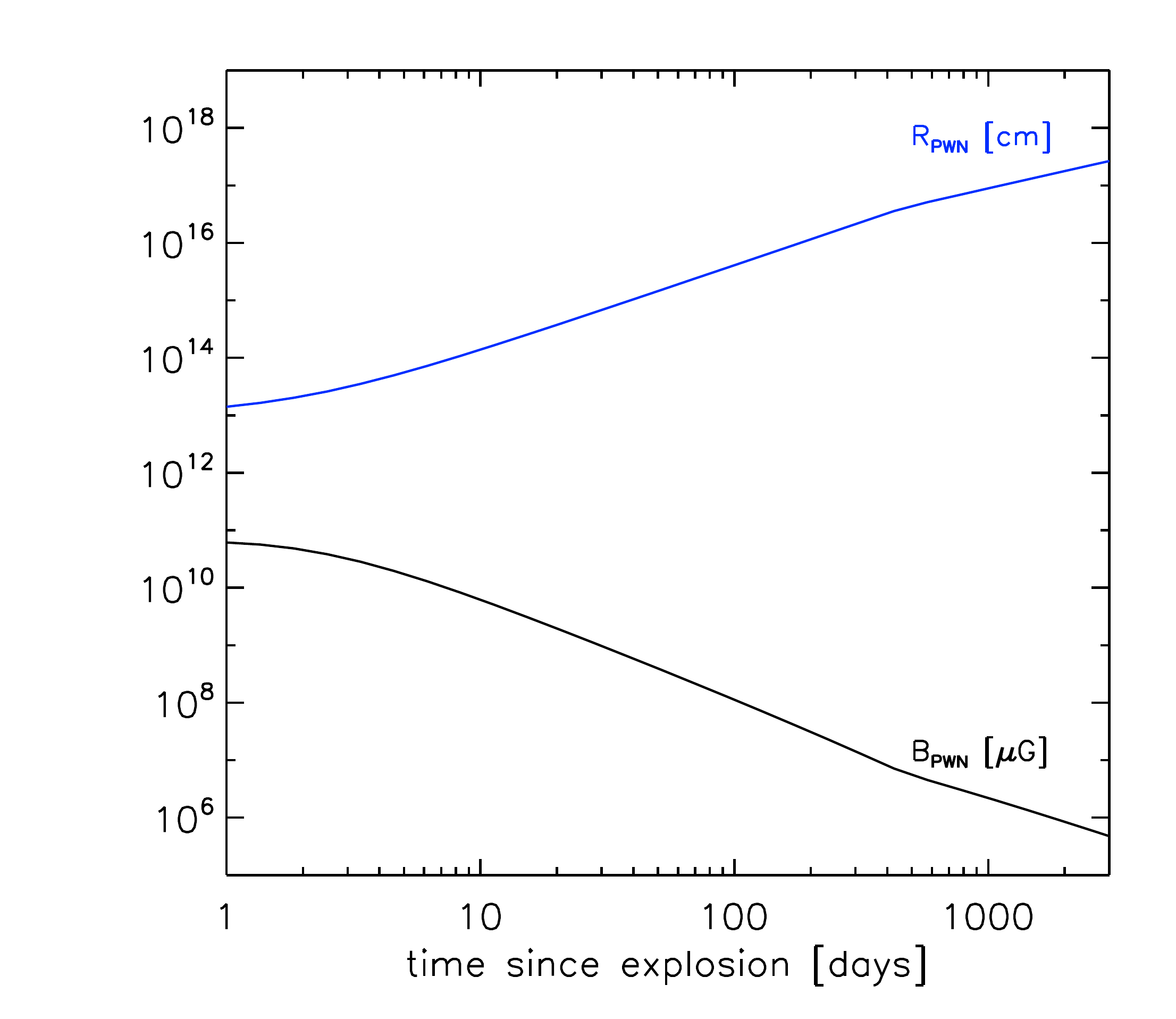} 
\caption{Evolution in time of the radius, $R_{\rm PWN}$, and the magnetic field strength, $B_{\rm PWN}$, of a pulsar wind nebula, assuming no shell disruption (Eqs.~\ref{eq:Rpwn_noshelldisr} and \ref{eq:Rpwn_tp}) and $\eta_{\rm B}=0.01$, calculated for a SN ejecta with $M_{\rm ej} = 5\,M_\odot$ and $E_{\rm ej}=10^{51}\,$erg\,s$^{-1}$, embedding a pulsar with dipole magnetic field of $B=10^{13}\,$G and period $P_{\rm i}=1\,$ms. } \label{fig:PWN}
\end{center}
\end{figure}

The magnetic field strength in the pulsar wind nebula can then be estimated assuming a fraction of magnetization $\eta_{\rm B}$ of the luminosity injected by the wind (see Fig.~\ref{fig:PWN})
\begin{equation}
B_{\rm PWN} = \left(8\pi \eta_{\rm B} \int_0^t L_{\rm p}(t'){\rm d}t' \right)^{1/2}R_{\rm {PWN}}(t)^{-3/2}\ .
\end{equation}
The value of $\eta_B$ could vary between $0.01-1$, according to pulsar wind nebulae. 

\section{Bolometric radiation}\label{section:lightcurves}

In what follows, we calculate the total radiation expected from the supernova ejecta+pulsar wind nebula. The evolution of the ejecta is computed assuming a one zone core-collapse model. This approximation is debatable for times $t\lesssim t_{\rm d}$, as the radiation should be mainly emitted in the central regions, close to the pulsar wind nebula, and not uniformly distributed as the matter over a single shell. This is not expected to be limiting for our study however, as we are most interested in the late-time light curves (a few years after the explosion), when the ejecta starts to become optically thin. 

How much energy of the pulsar wind will be transformed into radiation depends on many factors such as the nature of the wind (leptonic, hadronic or Poynting flux dominated), the efficiency of particle acceleration and of radiative processes. In a first step, these conditions can be parametrized by setting a fraction $\eta_\gamma$ of the wind energy $E_{\rm p}$ that is converted to radiative energy (thermal or non thermal) in the pulsar wind nebula.\\

Under the one zone model approximation, the radiation pressure dominates throughout the remnant, $P=E_{\rm int}/3V$, with $V$ the volume of the ejecta. The internal energy then follows the law:
\begin{equation}\label{eq:Eintlaw}
\frac{1}{t}\frac{\partial}{\partial t}[E_{\rm int}t] = \eta_\gamma L_{\rm p}(t) -L_{\rm rad}(t)\ .
\end{equation}
The radiated luminosity $L_{\rm rad}$ depends on the ejecta optical depth: 
\begin{eqnarray}
\frac{L_{\rm rad}(t)}{4\pi R^2} &=&\frac{E_{\rm int}c}{(4\pi/3)R^3}\quad t>t_{\rm thin}\\
&=&\frac{E_{\rm int}c}{(4\pi/3)\tau R^3} \quad t\le t_{\rm thin}
\end{eqnarray}
which yields
\begin{eqnarray}\label{eq:Lrad}
L_{\rm rad}(t) &=&\frac{3}{\beta_{\rm ej}}\frac{E_{\rm int}}t{}\quad t>t_{\rm thin}\\
&=&\frac{E_{\rm int}t}{t_{\rm d}^2} \quad t\le t_{\rm thin}
\end{eqnarray}
where we note $\beta_{\rm ej}\equiv v_{\rm ej}/c$.
For $t<t_{\rm thin}$, we assumed that the totality of the luminosity $\eta_\gamma L_{\rm p}$ deposited in the ejecta as photons is thermalized, and used the diffusion transport approximation \cite{Arnett80}. In the optically thin regime, photons do not diffuse and propagate straight out of the ejecta.

\begin{figure*}
\begin{center}
\includegraphics[width=0.49\textwidth]{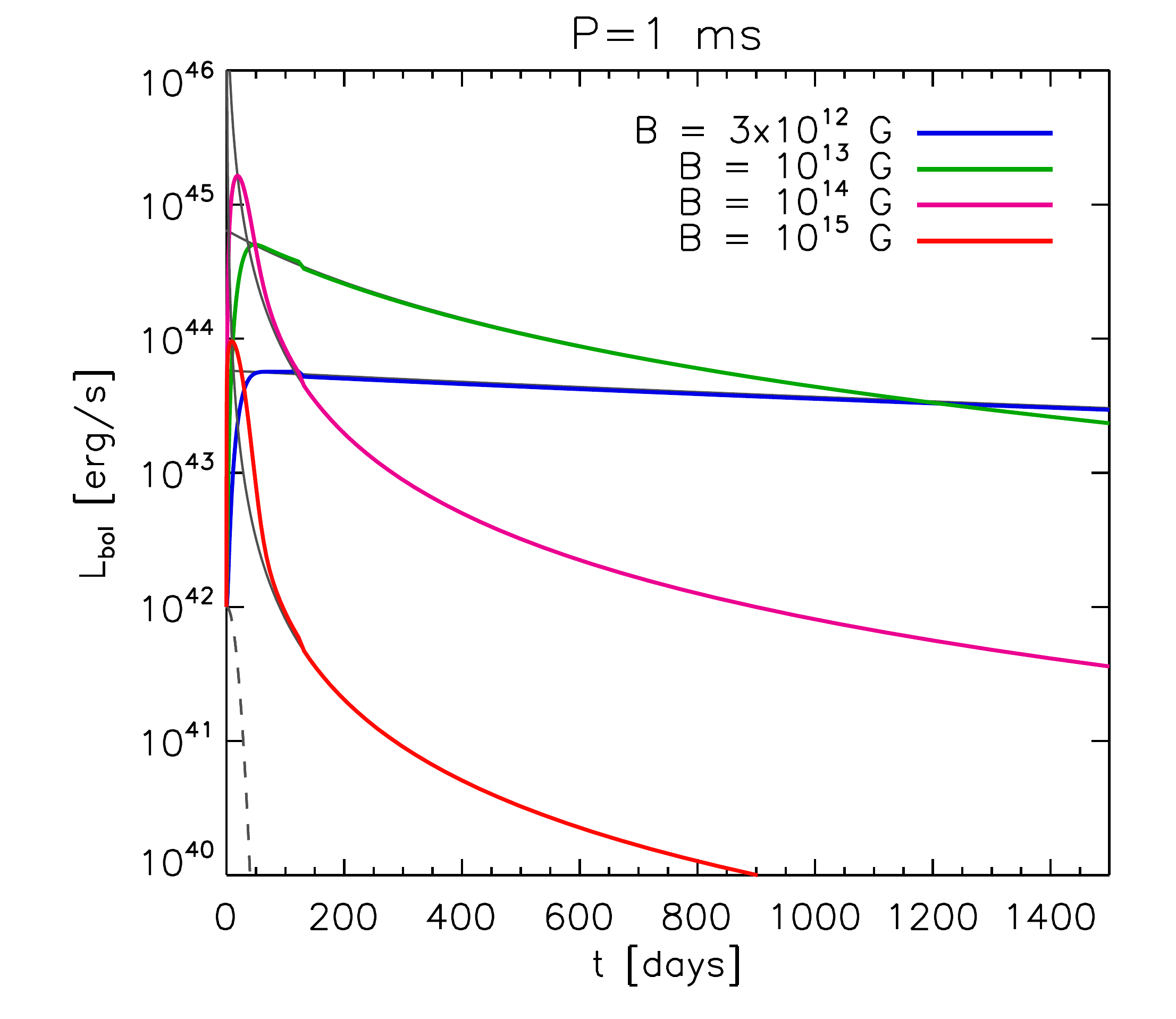} 
\includegraphics[width=0.49\textwidth]{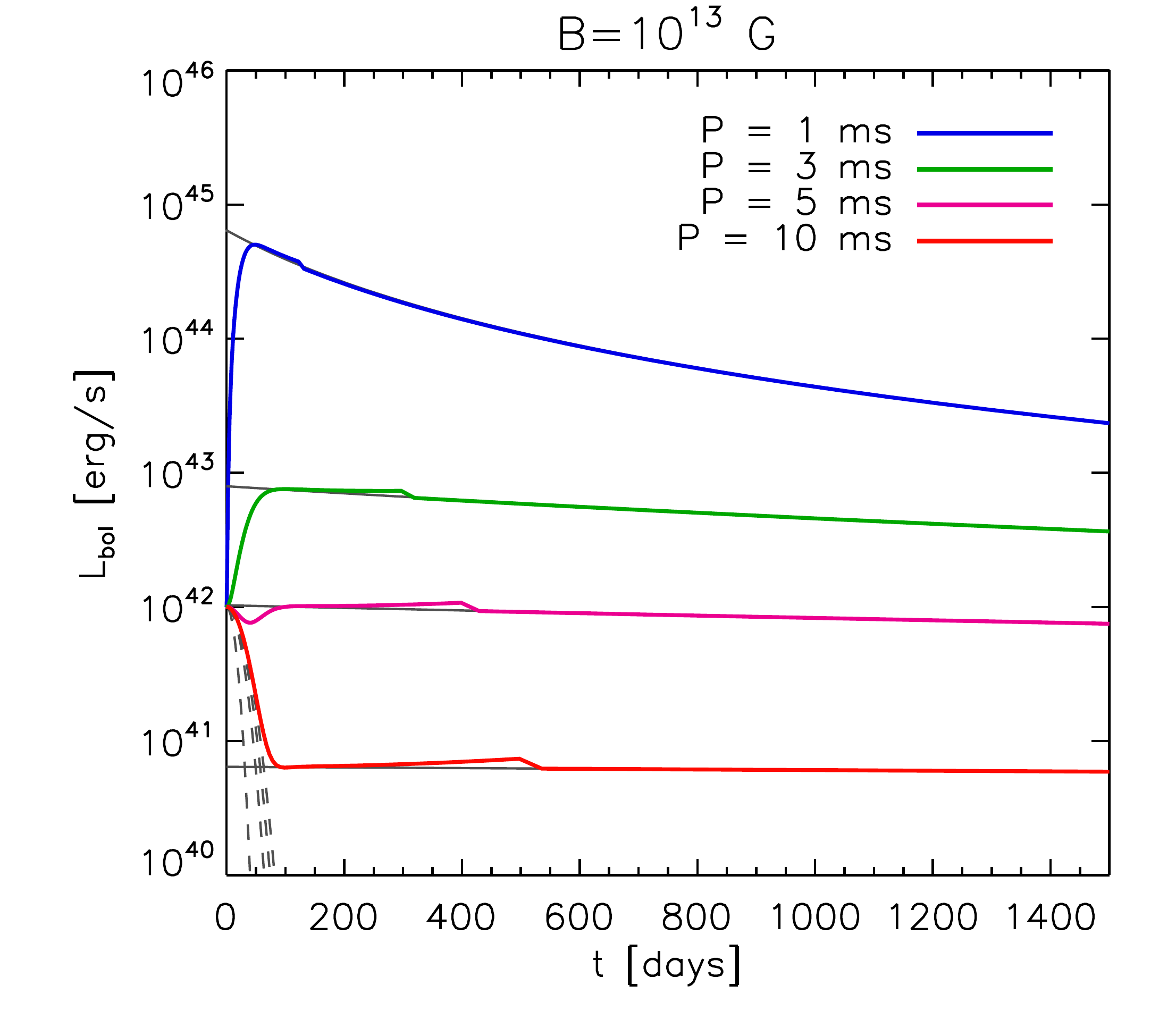} 
\caption{Evolution of the bolometric radiated luminosity of the supernova as a function of time. The pulsar has a dipole magnetic field of increasing strength as indicated, and period $P_{\rm i}=1$\,ms (left panel), and $B=10^{13}\,$G and increasing periods as indicated in the legend (right panel). The supernova ejecta has $M_{\rm ej} = 5\,M_\odot$ and $E_{\rm ej}=10^{51}\,$erg. The gray lines give the evolution of the pulsar luminosity $L_{\rm p}$ for each initial spin period. The gray dashed lines are the contribution of the ordinary core-collapse supernova to the radiated luminosity $L_{\rm SN}$. We have assumed $\eta_\gamma=1$.  The slight discontinuity is due to the numerical calculation of the integral in Eq.~(\ref{eq:Eint}), and marks the transition between $t<t_{\rm thin}$ and $t<t_{\rm thin}$.}\label{fig:Lrad}
\end{center}
\end{figure*}

Equation~(\ref{eq:Eintlaw}) yields
\begin{eqnarray}
E_{\rm int}(t) &=& \frac{\eta_\gamma E_{\rm p}}{1+3/\beta_{\rm ej}}\left[h_1\left(\frac{t}{t_{\rm p}}\right)-h_2\left(\frac{t}{t_{\rm p}}\right)\right]   \quad t>t_{\rm thin}\\
&=& \frac{1}{t} \,e^{-\frac{t^2}{2t_{\rm d}^2}}\, \left[ \int_{t_{\rm ex}}^t \,e^{\frac{x^2}{2t_{\rm d}^2}}\, \frac{\eta_\gamma E_{\rm p}t_{\rm p}x}{(t_{\rm p}+x)^2}{\rm d}x + E_{\rm ej}\,t_{\rm ex}\,e^{\frac{t_{\rm ex}^2}{2t_{\rm d}^2}} \right] \nonumber\\
&&\qquad  t\le t_{\rm thin}\ .\label{eq:Eint}
\end{eqnarray}
The hypergeometric functions are noted:
\begin{eqnarray}
h_1(x) &\equiv& \,_2F_1(1,1+3/\beta_{\rm ej},2+3/\beta_{\rm ej},-x)\\
h_2(x) &\equiv& \,_2F_1(2,1+3\beta_{\rm ej},2+3/\beta_{\rm ej},-x)\, .
\end{eqnarray}
Note that $L_{\rm rad}(t)\sim \eta_\gamma L_{\rm p}(t)$ for $t>t_{\rm thin}$.

To calculate the total bolometric radiated luminosity, we add to $L_{\rm rad}(t)$ the contribution of the ordinary core-collapse supernova radiation $L_{\rm SN}(t)$. $L_{\rm SN}(t)$ is calculated following Eq.~(5) of Chatzopoulos et al. (2012), assuming an initial luminosity output of $10^{42}\,$erg/s, as is estimated by Woosley \& Heger (2002) in their Eq.~(41), for $M_{\rm ej}=5\,M_\odot$ and $E_{\rm ej}=10^{51}\,$erg\,s$^{-1}$. $L_{\rm SN}$ only contributes when $E_{\rm p}<E_{\rm SN}$. \\

Fig.~\ref{fig:Lrad} presents the bolometric luminosity radiated from the ejecta+PWN system for various sets of pulsar parameters. Again, the \cite{Arnett80} approximation is not necessarily valid for $t<t_{\rm d}$, where the radiation should not be distributed over the whole ejecta. Even with a $\eta_{\gamma}<10\%$, the plateau in the light curve a few years after the explosion is highly luminous, especially for $P=1\,$ms. This high luminosity plateau stems from the injection of the bulk of the pulsar rotational energy a few years after the supernova explosion. The luminosity is quickly suppressed for high $B$ (for magnetar-type objects), due to the fast spin-down. Supernovae embedding isolated millisecond pulsars with standard magnetic field strengths would thus present unique radiative features observable a few years after their birth. 

Note however that the luminosity represented here is the bolometric one. The emission should shift from quasi-thermal to high energy after a few years, depending on the evolution of the opacity of the ejecta. The emission at different energies is discussed in the next section.

\section{Thermal/non-thermal emissions}\label{section:thermal}

The bolometric radiation calculated in the previous section stems from the re-processing of high energy radiation created at the base of the SN ejecta, in the PWN region. In the standard picture of PWN, high energy particles (leptons and hadrons) are injected at the interface between the pulsar wind and the ejecta, and radiate high energy photons (X rays and gamma rays). 
These high energy photons can be either thermalized if the medium (the PWN and/or the SN ejecta) is optically thick to these wavelengths, or can escape from the ejecta and be observed as a high energy emission, if the medium they have to propagate through is optically thin. 
In this section, we calculate in more detail the emission a few years after the explosion, concentrating mainly on the case of a leptonic wind. \\

Upstream of the termination shock, the energy of the pulsar wind is distributed between electrons and positrons, ions and magnetic fields. The fraction of energy imparted to particles is not certain, especially at these early times. Near the neutron star, the Poynting flux is likely to be the dominant component of the outflow energy. After many hundreds of years, observational evidence show that the energy repartition at the termination shock of pulsar wind nebulae is dominated by particles (e.g., Arons 2008). 
The conventional picture is thus that all but $\sim 0.3-1\%$ of the Poynting flux has already been converted into the plasma kinetic energy by the time the flow arrives the termination shock \citep{Kennel84a,Kennel84b,Emmering87,Begelman92}, $\sim 1\%$ appearing to be a level required to reproduce the observed shape of the Crab Nebula \citep{Komissarov04,DelZanna04}. How this transfer happens is subject to debate (see, e.g., \citealp{Kirk09}).

Particles and the Poynting flux are injected in the pulsar wind nebula at the termination shock. 
We will note the energy repartition between electrons and positrons, ions and the magnetic field in the pulsar wind nebula: $L_{\rm p}=(\eta_{\rm e}+\eta_{\rm i}+\eta_{\rm B})L_{\rm p}$. The ratio between $\eta_{\rm i}$ and $\eta_{\rm e}$ is the subject of another  debate (see e.g., \citealp{Kirk09}). However, various authors (e.g., \citealp{Gelfand09,Fang10,Bucciantini11,Tanaka11}) seem to fit satisfactorily the observed emissions for various late time pulsar wind nebulae without adding any hadronic injection. We will thus focus on the emission produced for winds dominated by a leptonic component at the termination shock. 

Note that if protons are energetically dominant in the wind, \cite{Amato03} calculated that a large flux of neutrinos, gamma-rays and secondary pairs from p-p pion production should be expected from Crab-like pulsar wind nebulae around a few years after the supernova explosion. They estimate that the synchrotron emission from secondaries will be negligible, while TeV photon and neutrino emission could be detectable by current instruments if such young objects were present in our Galaxy. 

Only 1\% of the relativistic ions and magnetic fields components of the wind can be converted into thermal energy in the ejecta (\citealp{Chevalier77}). This fraction can be amplified in presence of, e.g., Rayleigh-Taylor mixing, or high energy cosmic ray diffusion into the ejecta.

\subsection{Pair injection in the PWN}

According to the original idea by \cite{Kennel84a}, the pair injection spectrum into the pulsar wind nebula should present a Maxwellian distribution due to the transformation of the bulk kinetic energy of the wind into thermal energy, and a 
non-thermal power-law tail formed by pairs accelerated at the shock.
Hybrid and PIC simulations have shown indeed such a behavior (e.g., \citealp{Bennett95,Dieckmann09,Spitkovsky08}). \cite{Spitkovsky08} finds that 1\% of the particles are present in this tail, with 10\% of the total injected energy. The bulk of the particle energy would then be concentrated around the kinetic energy upstream of the termination shock
\begin{eqnarray}\label{eq:epsc}
\epsilon_{\rm b} &=& kT_{\rm e} = \gamma_{\rm w} m_{\rm e} c^2\ \\
&\sim& 5\times 10^{11}\,{\rm eV}\, \frac{\gamma_{\rm w}}{10^6}\ ,
\end{eqnarray}
with $\gamma_{\rm w}$ the Lorentz factor of the wind.
The non-thermal tail would start around $\epsilon_{\rm b}$ and continue up to higher energies with a spectral index $\gtrsim 2$.  In practice, from a theoretical point of view, Lorentz factors as high as $\gamma_{\rm w}\sim 10^6$ are difficult to reach, and current simulations are only capable of producing $\gamma_{\rm w}$ of order a few hundreds \citep{Spitkovsky08,Sironi09}. 

However, observationally, various authors (\citealp{Kennel84a}, but also more recently, e.g., \citealp{Gelfand09,Fang10,Bucciantini11,Tanaka11}) demonstrated that the non-thermal radiation produced by the injection of either one single power-law or a broken power-law peaking around $\epsilon_{\rm b}\sim 1\,$TeV, and extending up to PeV energies, could fit successfully the observed emission of various young pulsar wind nebulae. Such a high break energy implies either a high Lorentz factor for the wind $\gamma_{\rm w}\sim 10^{5-6}$, or an efficient acceleration mechanism enabling particles to reach $0.1-1$\,TeV energies. At higher energies, another acceleration mechanism has to be invoked to produce particles up to PeV energies. 
\cite{Bucciantini11} discuss that $\epsilon_{\rm b}$ could possibly be viewed as a transition energy between Type II and Type I Fermi acceleration from low at high energies. This would provide a physical explanation to the broken power-law shape, and alleviate the issue of the high wind Lorentz factor. 
At high energies, acceleration could also happen in the course of reconnection of the striped magnetic field in the wind, at the termination shock \citep{Lyubarsky03,Petri07}. However, it is not clear yet whether this process can lead to a non-thermal particle distribution. 
One can expect additional particle acceleration in the wind itself, via surf-riding acceleration \citep{Chen02,Contopoulos02,Arons02,Arons03}. This non-thermal component would not necessarily be processed when injected at the shocks if the particle Larmor radii are large compared to the size of the shock. 

In the following, we will assume that pairs are injected in the pulsar wind nebula following a broken power-law of the form 
\begin{equation}\label{eq:Neinj}
\frac{{\rm d}\dot{N}}{{\rm d}\epsilon}(\epsilon,t) = \frac{\eta_{\rm e}L_{\rm p}(t)}{\epsilon_{\rm b}^2} \,\left\{
\begin{array}{ll}
(\epsilon/\epsilon_{\rm b})^{-\alpha}&\quad \mbox{if} \quad \epsilon_{\rm min}\le \epsilon < \epsilon_{\rm b} \\
(\epsilon/\epsilon_{\rm b})^{-\beta}&\quad \mbox{if} \quad  \epsilon_{\rm b}\le \epsilon \le \epsilon_{\rm max}  
\end{array}\right. 
\end{equation}
where $\alpha<2<\beta$, $\epsilon_{\rm min}$ and $\epsilon_{\rm max}$ are the minimum and maximum cut-off energies respectively, and $\epsilon_{\rm b}$ the peak of the injection distribution $\epsilon({\rm d}N/{\rm d}\epsilon)\sim 0.1-1$\,TeV.  It is commonly assumed that $\epsilon_{\rm b} \propto\gamma_{\rm w}\propto \sqrt{L_{\rm p}(t)}$, but such an assumption would imply very high wind Lorentz factors ($>10^9$) at early times, that seem incompatible with the simulations and theoretical models discussed above. It is likely that the Lorentz factor experiences a saturation above a certain value, and for simplicity, we will assume that $\epsilon_{\rm b}$ is constant over time. For our purpose of deriving a rough estimate of the fraction of high energy emission that can escape the ejecta at early times, such an approximation will suffice. A thorough calculation of the emission spectrum would require time-dependent energy loss calculations for particles beyond the one zone approximation that we use here. \cite{DelZanna04,DelZanna06} have shown indeed that the high energy emission is strongly affected by the details of the flow dynamics just downstream of the termination shock.

\begin{figure}
\begin{center}
\includegraphics[width=\columnwidth]{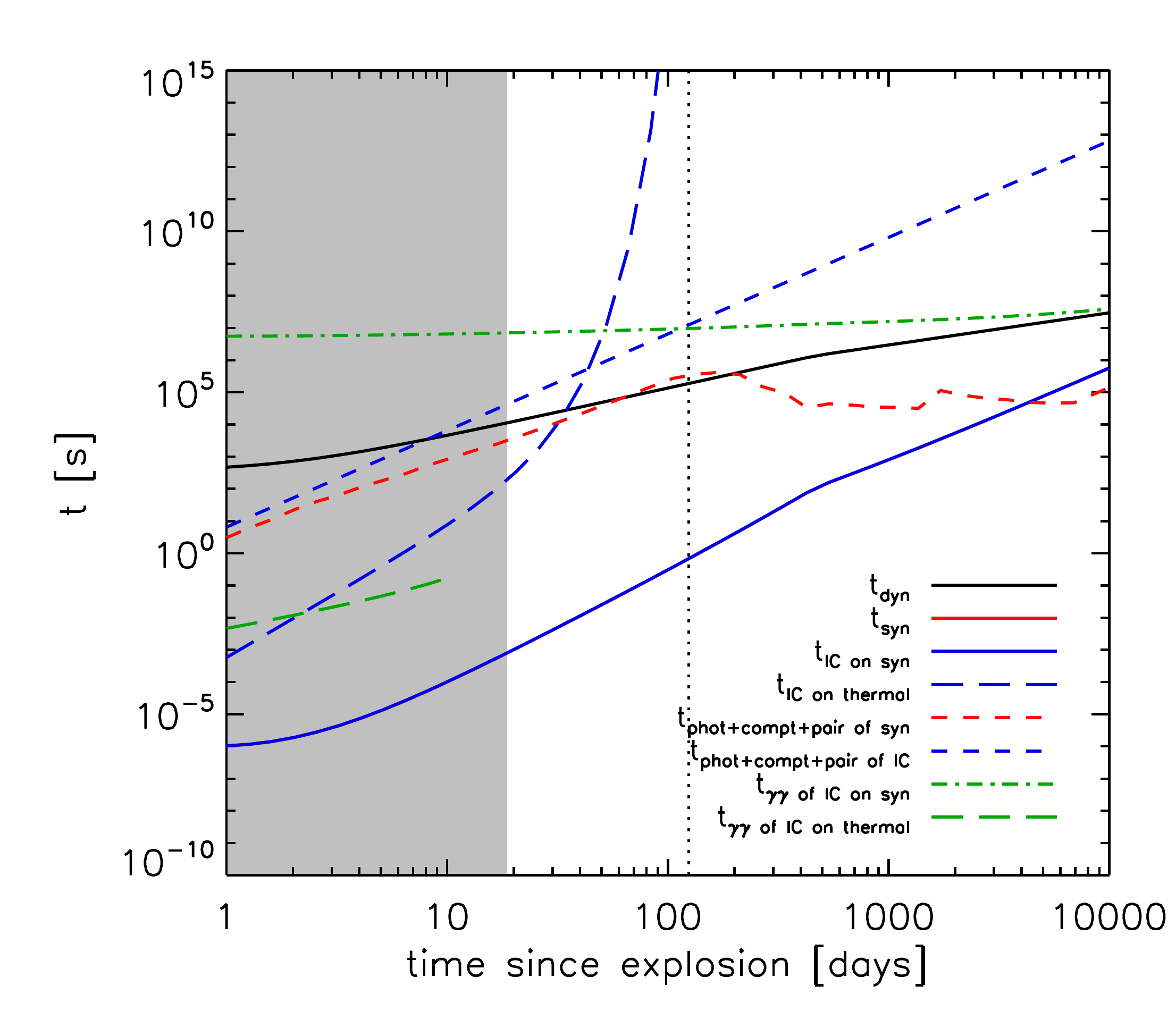} 
\caption{Timescales at play in the radiation emission of a PWN, for the same system as in Fig.~\ref{fig:PWN}, assuming an electron injection break energy $\epsilon_{\rm b}=0.1\,$TeV, and $\eta_B=0.01$. The dynamical timescale, $t_{\rm dyn}$ (black solid), and cooling timescales via synchrotron, $t_{\rm syn}$ (red solid), via self-compton, $t_{\rm IC, syn}$ (blue solid), and via IC off thermal photons (blue long dashed), are compared to thermalization timescales. The thermalization timescales via photoelectric absorption, Compton scattering, and pair production are in dashed lines (in red for the synchrotron photons and blue for the IC photons). Absorption by $\gamma-\gamma$ interaction at high energies are in green lines (dot-dashed for the interaction between IC photons on synchrotron photons, and long dashed for the interaction of IC photons on the thermal photons of the SN ejecta).
The gray shaded region corresponds to $t<t_{\rm d}$, where the \citet{Arnett80} approximation is not valid. And the dotted line indicates $t=t_{\rm thin}$. }\label{fig:times}
\end{center}
\end{figure}

\begin{figure*}
\begin{center}
\includegraphics[width=0.49\textwidth]{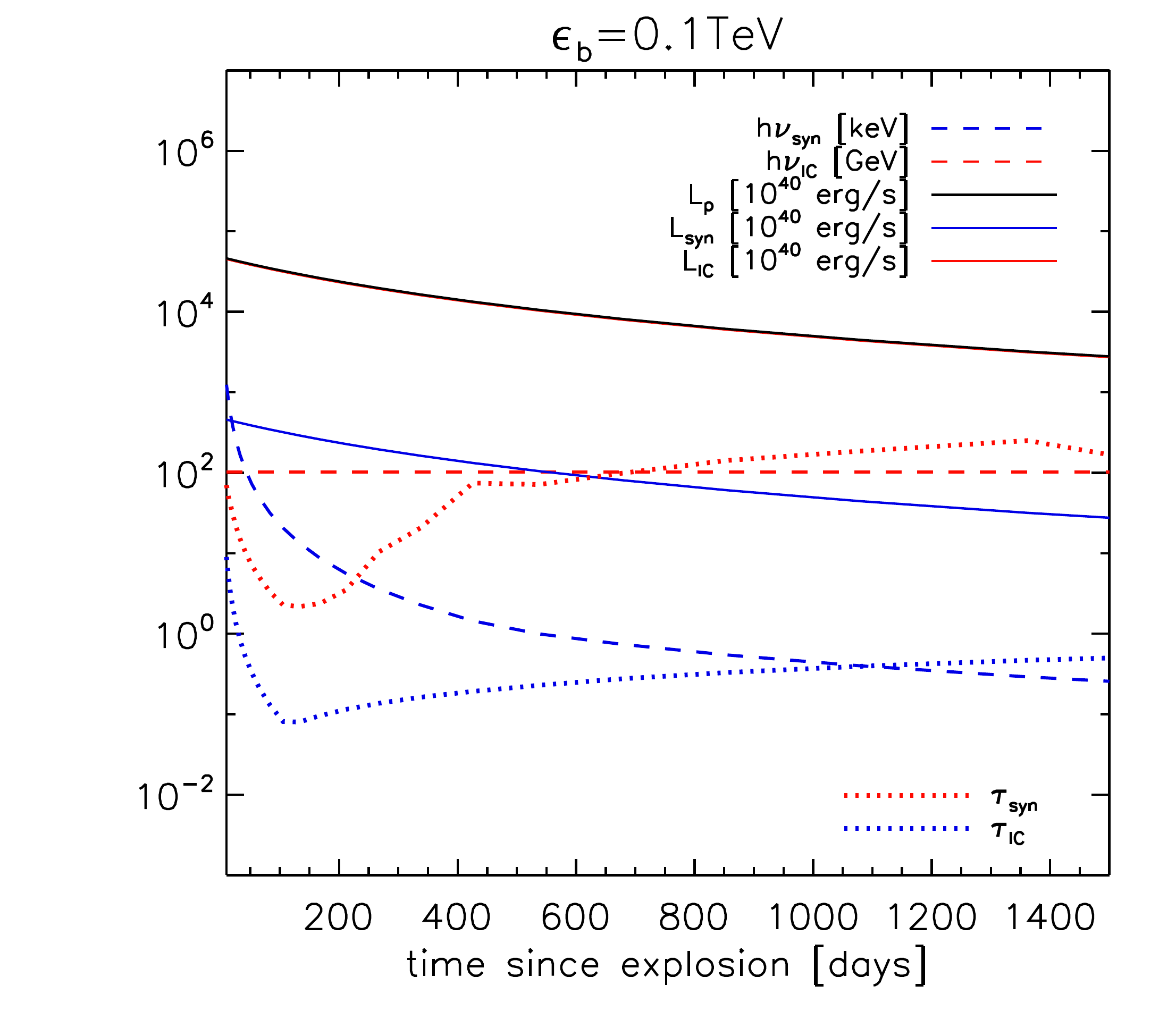} 
\includegraphics[width=0.49\textwidth]{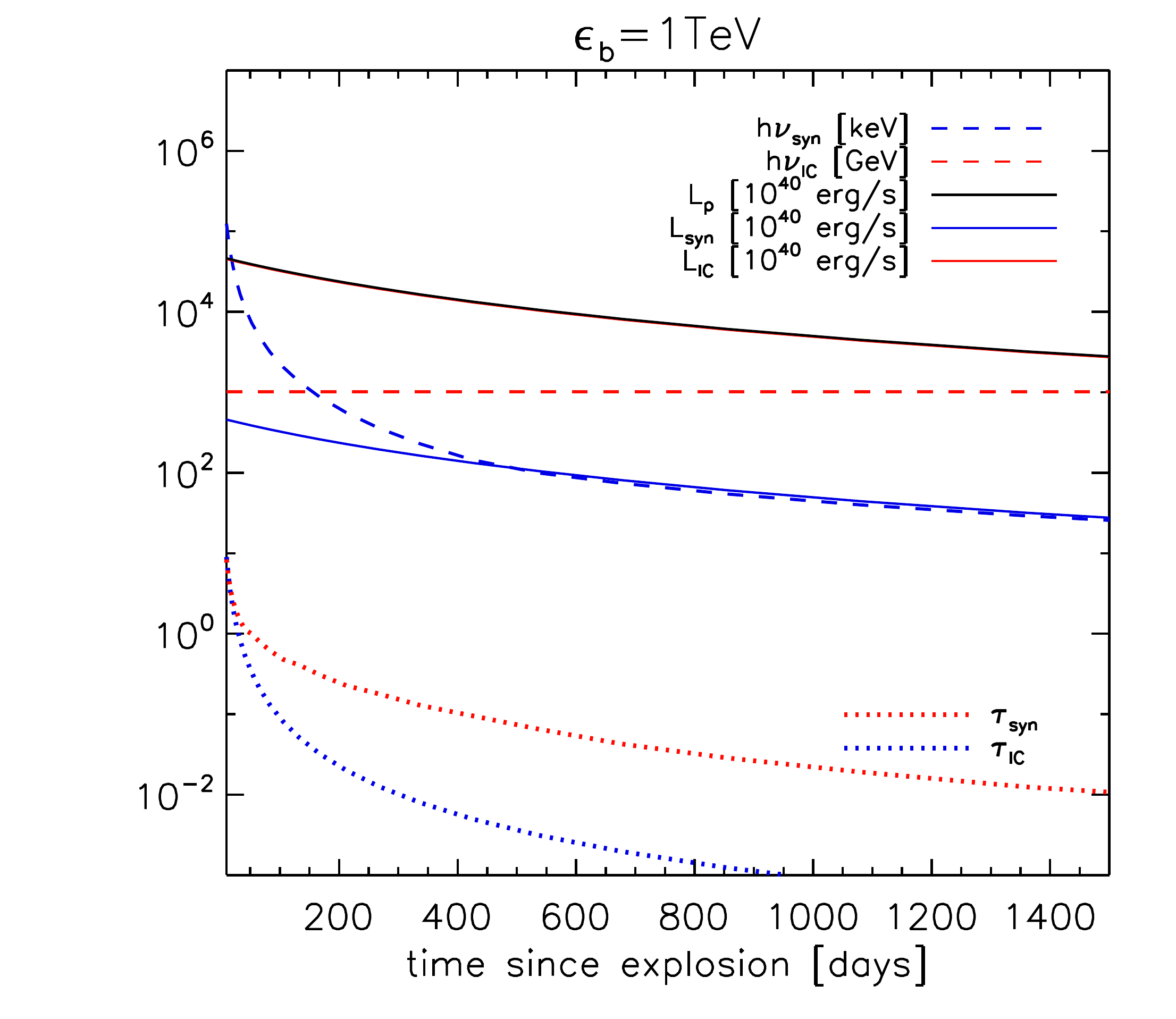} 
\caption{Evolution in time of radiative quantities of the same system as in Fig.~\ref{fig:PWN}, with $\eta_{B}=0.01$, $\eta_e=1-\eta_B$, and for a particle injection break energy of $\epsilon_{\rm b}=0.1\,$TeV (left) and $\epsilon_{\rm b}=1\,$TeV (right). Blue dashed line: characteristic energy of synchrotron radiation; red dashed lines: characteristic energy of photons produced via self-comptonization of synchrotron photons. Thin black solid line: luminosity injected by the pulsar. Red and blue solid lines: luminosity emitted in synchrotron and IC radiations. Dotted blue line: optical depth of ejecta of composition 60\% H, 30\% He and 10\% C, to photons of energy $h\nu_{\rm c}$. Dotted red lines: optical depth of ejecta to photons of energy $\nu_{\rm IC}$.} \label{fig:rad}
\end{center}
\end{figure*}

\subsection{Radiation by accelerated pairs}

The bulk of the electron distribution will predominantly radiate in synchrotron and experience inverse Compton (IC) scattering off the produced synchrotron photons. The cooling timescales of these processes, as well as the dynamical timescale $t_{\rm dyn}=R_{\rm PWN}/c$ of the PWN are indicated in Fig.~\ref{fig:times}.  Inverse Compton scattering off the thermal photons of the ejecta and off the Cosmic Microwave Background (CMB) are negligible compared to the former two processes. Figure~\ref{fig:times} also demonstrates that the cooling timescale of IC scattering off the thermal photons of the SN ejecta is much longer than the timescale for self-comptonization of the synchrotron emission. This estimate includes only the contribution of the thermal photons of the standard supernova ejecta, as the thermalization of the non-thermal components described here happen on larger timescales, in the optically thin regime which is of interest to us.

The synchrotron cooling timescale of accelerated electron reads
\begin{equation}
t_{\rm syn} = \frac{3 m_e^2c^3}{4\sigma_{\rm T}\epsilon_{\rm c}U_B}\  ,
\end{equation}
with $U_B=B_{\rm PWN}^2/8\pi$.
At the early stages that we consider here, the characteristic energy of radiating particles is  $\epsilon_{\rm c}(t)=\epsilon_{\rm b}$.
The characteristic synchrotron radiation frequency can be expressed 
\begin{equation}
\nu_{\rm c}(t) = 0.29\frac{3e B_{\rm PWN}(t)}{4\pi m_e^3 c^5}\epsilon_{\rm c}^2\ .\label{eq:nu_c}
\end{equation}
Accelerated electrons also scatter off these synchrotron photons by IC, producing photons at energy
\begin{eqnarray}
\nu_{\rm IC} &=& \nu_{\rm KN}\equiv\frac{m_e c^2}{\gamma_e h} \quad \mbox{if } \nu_{\rm c}>\nu_{\rm KN}\ ,\\
&=& \nu_{\rm c}\quad \mbox{if } \nu_{\rm c}\le \nu_{\rm KN}\ ,
\end{eqnarray}
with a cooling timescale 
\begin{equation}
t_{\rm IC, syn} = \frac{3m_e^2c^4}{4\sigma_{\rm T}c\epsilon_{\rm c}U_{\rm syn}}\ ,
\end{equation}
where $U_{\rm syn}$ is the synchrotron photon energy density.
Electrons radiate in synchrotron and self-compton processes with the following power ratio: $P_{\rm IC}/P_{\rm syn}=U_{\rm syn}/U_B$. Assuming that the energy of the accelerated electron population is concentrated in its peak energy $\epsilon_{\rm b}$, this implies synchrotron and IC luminosities of $L_{\rm syn}=\eta_B\eta_{e}/(\eta_B+\eta_{e})L_{\rm p}$ and $L_{\rm IC}=\eta_{e}^2/(\eta_B+\eta_{e})L_{\rm p}$ respectively. Obviously, the value of $\eta_B$ has an impact on the synchrotron emission, but not on the IC emission.

Figure~\ref{fig:rad} presents the evolution in time of the luminosities $L_{\rm p}$, $L_{\rm syn}$, and $L_{\rm IC}$, as well as the emission frequencies $\nu_{\rm c}$ and $\nu_{\rm IC}$. The IC radiation is mostly emitted at the break energy of the injection of electrons. The synchrotron emission spans from gamma/X-ray (until a few years) to optical wavelengths (after thousands of years). At the time of interest in this study, X-rays are thus mainly emitted between $0.1-100$\,keV for $\epsilon_{\rm b}=0.1\,$TeV, and around $100\,{\rm keV}-1\,{\rm GeV}$ for $\epsilon_{\rm b}=1\,$TeV.

\subsection{Thermalization in the ejecta}

\begin{figure}
\begin{center}
\includegraphics[width=\columnwidth]{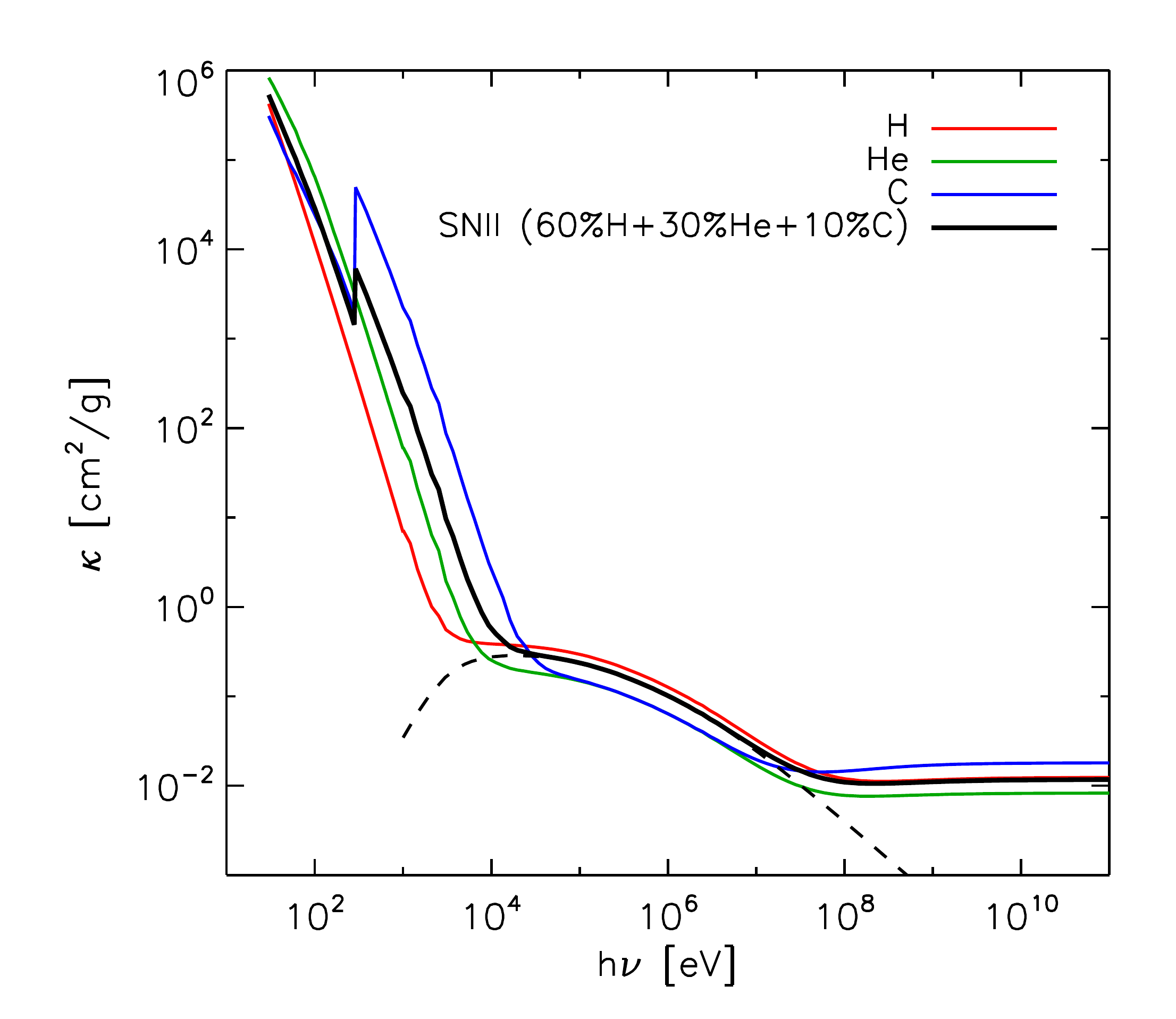} 
\caption{Atomic scattering opacities of high energy photons on H, He, C, and a mixture of composition mimicking that of a type II core-collapse supernova ejecta (60\% H, 30\% He and 10\% C). The black dashed line indicates the contribution of Compton scattering in the latter composition case. From http://henke.lbl.gov/optical\_constants/ for $30\,{\rm eV}\le h\nu <1\,{\rm keV}$ and http://physics.nist.gov/cgi-bin/Xcom/ for $1\,{\rm keV}\le h\nu <100\,{\rm GeV}$. }\label{fig:kappa}
\end{center}
\end{figure}

The X-ray opacity ($\sim 0.1-100\,$keV) is dominated by photoelectric absorption in metals. Above $\sim 100\,$keV, very hard X-rays and gamma-rays experience predominantly Compton scattering, and pair production above $\sim 10\,$MeV. The opacities of these processes for various atomic media are given in Fig.~\ref{fig:kappa}. At a given time $t$, the optical depth of the ejecta to the characteristic synchrotron photon emission $\nu_{\rm c}$ reads
\begin{equation}
\tau_{\rm syn}(t) = v_{\rm ej} t\kappa_{\nu_{\rm c}}(t)\rho(t)\ .
\end{equation}
The dominant thermalization process for the TeV IC radiation is pair production by $\gamma-\gamma$ interactions (see Fig.~\ref{fig:times}). The timescale for thermalization via this process is however slightly longer than the dynamical time; hence the ejecta appears mostly optically thin to this radiation (see Fig.~\ref{fig:rad}). We note $\tau_{\rm IC}$ for the optical depth of the ejecta to the IC emission. 

The luminosity in the characteristic energies $h\nu_{\rm c}$ and $h\nu_{\rm IC}$ after thermalization, noted respectively $L_{\rm X}(t)$ and $L_{\gamma}(t)$, and the luminosity in thermal photons, $L_{\rm th}(t)$, are calculated as follows
\begin{eqnarray}
L_{\rm X}(t) &=& L_{\rm syn}(\nu_{\rm c},t)\,e^{-\tau_{\rm syn}(t)}\label{eq:Lxgamma}\\
L_{\gamma}(t) &=& L_{\rm IC}(\nu_{\rm IC},t)\,e^{-\tau_{\rm IC}(t)}\label{eq:Lxgamma}\\
L_{\rm th}(t) &=& L_{\rm rad}(t)-L_{\rm X}(t)-L_{\gamma}(t)\ .
\end{eqnarray}

\begin{figure*}
\begin{center}
\includegraphics[width=0.49\textwidth]{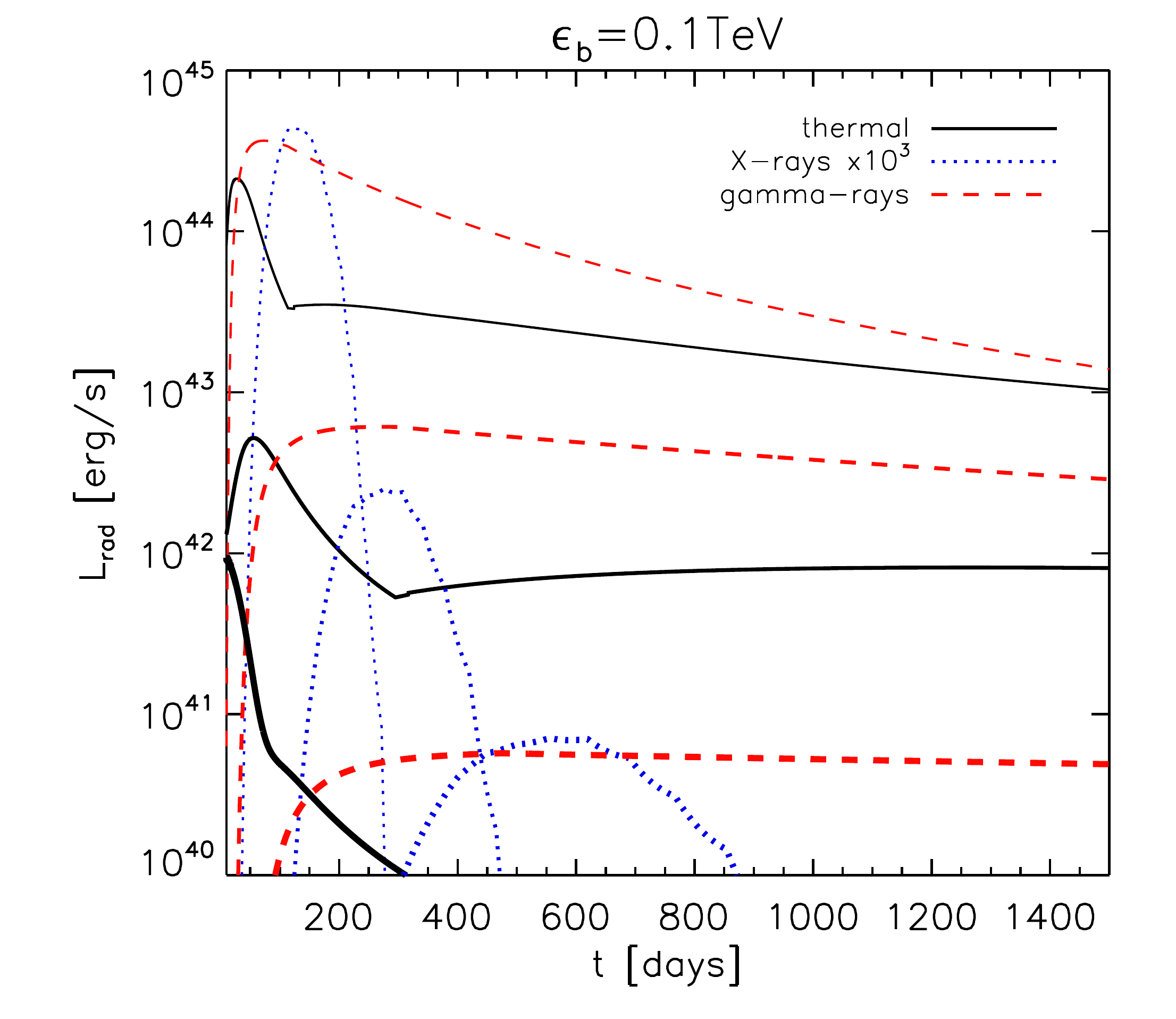} 
\includegraphics[width=0.49\textwidth]{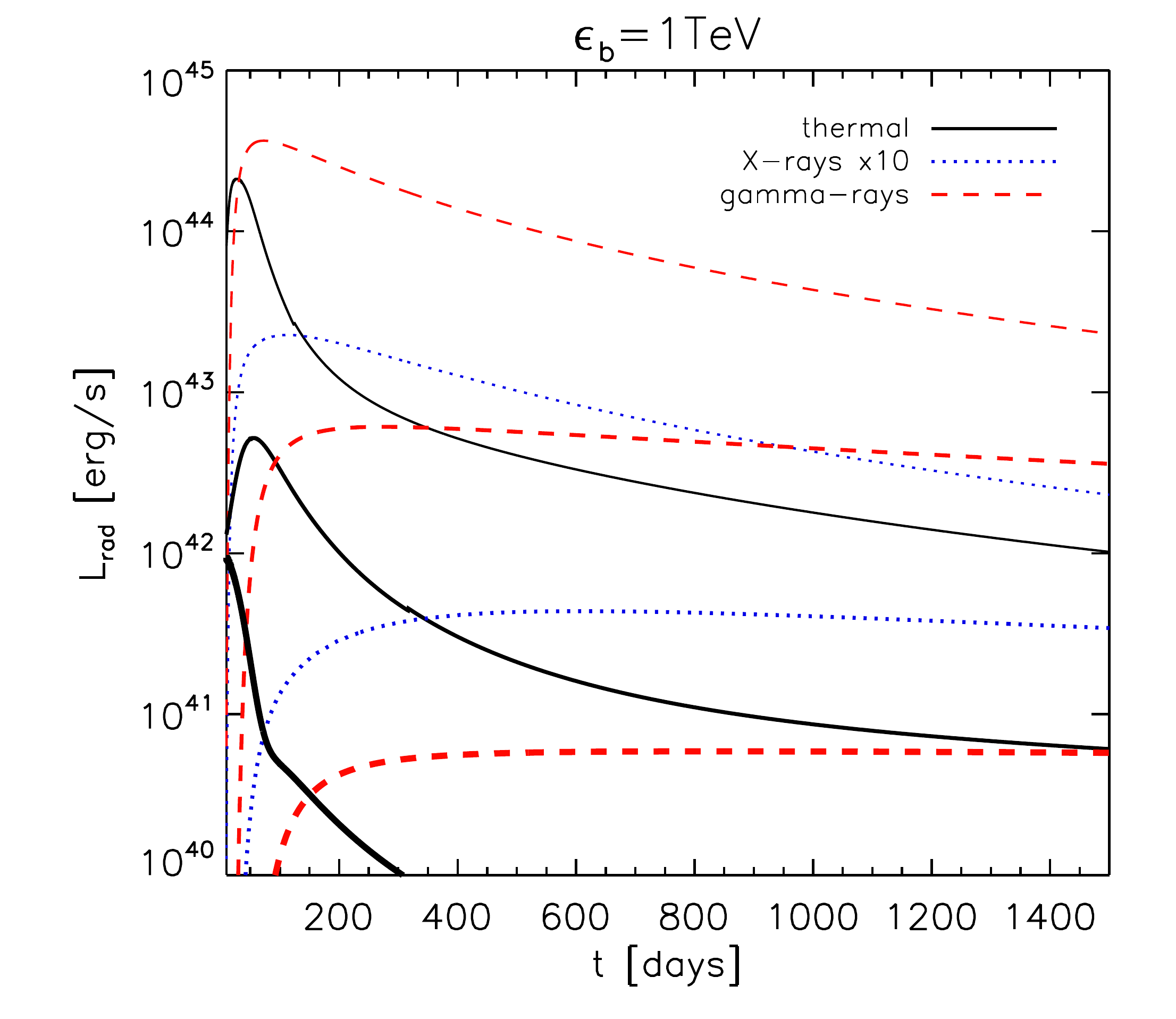} 
\caption{Thermal emission (black solid lines), non thermal X-ray emission (blue dotted) at energy indicated in Fig.~\ref{fig:rad} (emitted  mostly between $0.1-100$\,keV for $\epsilon_{\rm b}=0.1\,$TeV, and around $100\,{\rm keV}-1\,{\rm GeV}$ for $\epsilon_{\rm b}=1\,$TeV), and non thermal gamma-ray emission (red dashed) at 0.1\,TeV (left) and 1\,TeV (right), expected from a SN ejecta with $M_{\rm ej} = 5\,M_\odot$ and $E_{\rm ej}=10^{51}\,$erg\,s$^{-1}$, embedding a pulsar with dipole magnetic field of $B=10^{13}\,$G and period $P_{\rm i}=1, 3, 10\,$ms (increasing thickness), assuming $\eta_B=0.01$, $\eta_e=1-\eta_B$, and a break energy $\epsilon_{\rm b}=0.1\,$TeV (left) and $\epsilon_{\rm b}=1\,$TeV (right). Caution: for visibility, the X-ray luminosity is multiplied by $10^3$ on the left panel, and by 10 on the right panel. The slight discontinuities are numerical artifacts.}\label{fig:lumX}
\end{center}
\end{figure*}

Figure~\ref{fig:lumX} presents the thermal emission (black), X-ray emission (blue dotted) at $h\nu_{\rm c}\sim 0.1-100\,$keV for $\epsilon_{\rm b}=0.1\,$TeV (left panel) and $\sim 100\,{\rm keV}-1\,{\rm GeV}$ for $\epsilon_{\rm b}=1\,$TeV (right panel), and $0.1-1\,$TeV gamma ray emission (red dashed) expected from a SN ejecta with $M_{\rm ej} = 5\,M_\odot$ and $E_{\rm ej}=10^{51}\,$erg\,s$^{-1}$, embedding a pulsar with dipole magnetic field of $B=10^{13}\,$G and period $P_{\rm i}=1, 3, 10\,$ms (increasing thickness), assuming $\eta_B=0.01$, $\eta_e=1-\eta_B$, and a break energy $\epsilon_{\rm b}=0.1\,$TeV (left) and $\epsilon_{\rm b}=1\,$TeV (right). 

A decrease in flux is expected in the thermal component after a few months to years, when the ejecta becomes optically thin to gamma-rays. For a low break energy ($\epsilon_{\rm b}=0.1\,$TeV) the thermal component can then recover, as the X-ray emission vanishes, because of the increase of the ejecta optical depth for lower energy photons. 
One robust result is that, in both break energy cases, for fast pulsar rotation periods $P_{\rm i}\le 3\,$ms, the associated gamma-ray flux around $0.1-1\,$TeV emerges at a level that should be detectable at a few tens of Mpc, and remains strong over many years.

\section{Discussion, conclusion}\label{section:discussion}

We have estimated the thermal and non thermal radiations expected from supernova ejectas embedding pulsars born with millisecond periods, concentrating at times a few years after the onset of the explosion. The bolometric light curves should present a high luminosity plateau (that can reach $>10^{43}\,$erg/s) over a few years. A more detailed emission calculation considering the acceleration of leptons in the pulsar wind nebula region shows that an X-ray and a particularly bright TeV gamma-ray emission (of  magnitude comparable to the thermal peak) should appear around one year after the explosion. This non thermal emission would indicate the emergence of the pulsar wind nebula from the supernova ejecta.

The light curves calculated in this paper are simple estimates, that do not take into account second order effects of radioactive decay of $^{56}$Ni, recombination, etc. (see, e.g., \citealp{Kasen09}). The non-thermal components are also evaluated assuming that all the leptonic energy is concentrated in one energy bin. A more detailed analysis should be conducted, taking into account the shape of the spectra and its evolution in time, in order to get a more accurate representation of the emission, and for a thorough comparison with observational data. Depending on the spectral indices, a non monoenergetic electron injection spectrum could lead to a decrease of the peak luminosity of one order of magnitude. 

Our computation of the evolution of the PWN (radius, magnetic field) is also basic, and could benefit from more thorough estimations. Our toy model suffices however in the scope of this study, where the aim is to demonstrate the importance of multi-wavelength follow ups of SN lightcurves. We also assumed a relatively high magnetization $\eta_B$ of the wind at the termination shock, following estimates that reproduce the features of the Crab nebula \citep{Komissarov04,DelZanna04}. 

Several earlier works treat some of the aspects invoked above in more detail.  For example, in the context of evolution of pulsar wind nebulae, early works by \cite{Pacini73,Rees74,Bandiera84,Weiler80,Reynolds84} take into account the detailed evolution of particle energy distribution and radiation spectrum. Most of these works aim at calculating radiation features of observed plerions, a few hundred of years after the explosion. However, their modeling at earlier times, especially in the work of \cite{Reynolds84}, lays the ground for the more detailed calculations that should be performed in our framework.

The level of synchrotron emission predicted here can thus be viewed as optimistic values. However, the gamma-ray flux that is predicted does not depend on the magnetization, and remains fairly robust to most parameter changes.

Currently, only a handful of supernovae have been followed over a period longer than a year, and no object, except for SN 1987A, has been examined in X-rays or TeV gamma-rays a year after the explosion. Among the objects that have been followed in optical bands, SN 2003ma \citep{Rest11} has an abnormally bright luminosity at the peak, and a long bright tail over many years.  The six type II supernovae followed by \cite{Otsuka12} present various shapes of light curves, and a cut-off in the thermal emission after a few years. Our study demonstrates that the features in these light curves could also be due to the energy injection from the pulsar, that could compete with the other processes that are more commonly considered, such as the light echo of the peak luminosity, or the radioactive decay of $^{56}$Ni. An associated X-ray and TeV gamma-ray emission emerging around a few months to a year after the explosion would constitute a clear signature of pulsar rotational energy injection. It is also interesting to note that the emergence of a pulsar wind nebula has been recently reported from radio observations for SN 1986J \citep{Bietenholz10}, though over longer timescales than predicted for the objects studied in this paper.

Some authors \citep{Katz11,Svirski12} have discussed that shock breakouts from stars surrounded by a thick wind could lead to bright X-ray peak after a few months, similar to the signal discussed in this paper. This degeneracy can be overcome by the observation of the gamma-ray signal, which should be absent in the shock breakout scenario. Detailed analysis of the respective X-ray spectra should also help distinguish the two scenarios.

The follow up of bright type II supernovae over a few years after the explosion in different wave bands would thus reveal crucial information on the nature of the compact remnant. These suveys should be made possible with the advent of optical instruments such as LSST, and the use of powerful instruments for transient event detection, such as the Palomar Transient Factory or Pan-STARR. The bright X-ray signal should be detected by NuSTAR for supternovae out to redshifts $z\sim 0.5$, and the even brighter gamma-ray signal could be observed by HESS2, by the future Cerenkov Telescope Array (CTA), and by HAWK which will be the choice instrument to explore the transient sky at these energies. For CTA, an adequate survey of the sky outside the Galactic plane could spot gamma-ray sources of luminosity $10^{43}\,$erg/s as predicted by this work within a radius of $\sim 150\,$Mpc (G. Dubus, private comm.). Assuming a gaussian pulsar period distribution centered around $300\,$ms as in \cite{Faucher06}, implies that 0.3\% of the total population has spin periods $< 6\,$ms. With this estimate, one could find 4 bright sources within 150\,Mpc.
This is consistent with the numbers quoted in early works by \cite{Srinivasan84,Bhattacharya90}. These authors estimated the birthrate of Crab-like pulsar-driven supernova remnants to be of order 1 per 240 years in our Galaxy.

Pulsars born with millisecond periods embedded in standard core-collapse supernova ejectas, as described in this paper, are promising candidate sources for ultrahigh energy cosmic rays \citep{Fang12,Fang13}. In the framework of UHECRs, an injection of order 1\% of the Goldreich-Julian density into ions would suffice to account for the observed flux, assuming that 1\% of Type II supernovae give birth to pulsars with the right characteristics to produce UHECRs (i.e., pulsars born with millisecond periods and magnetic fields $B\sim 10^{12-13}\,$G, \citealp{Fang12}). The observation of the peculiar light curves predicted here could thus provide a signature for the production of UHECRs in these objects. Though no spatial correlation between arrival directions of UHECRs and these supernovae is expected (because of time delays induced by deflections in magnetic fields), an indication of the birth rate of these supernovae could already give direct constraints on this source model. Photo-disintegration and spallation of accelerated nuclei in the supernova ejecta could also lead to an abundant high energy neutrino production (\citealp{Murase09} consider such a neutrino production in the case of magnetars instead of fast-rotating pulsars), that could be detected with IceCube, and correlated with the position of identified peculiar supernovae.

\section*{Acknowledgments}

We thank L. Dessart, G. Dubus, M. Lemoine, K. Murase, E. Nakar, and J. Vink for very fruitful discussions. 
KK was supported at Caltech by a Sherman Fairchild Fellowship, and acknowledges financial support from PNHE. AO was supported by the NSF grant PHY-1068696 at the University of Chicago, and the Kavli Institute for Cosmological Physics through grant NSF PHY-1125897 and an endowment from the Kavli Foundation.

\bibliography{KPO}

\bsp

\label{lastpage}

\end{document}